\documentclass[traditabstract]{aa}  

\usepackage{graphicx}
\usepackage{txfonts}
\usepackage{ulem}
\usepackage{color}
\usepackage{longtable}
\usepackage{multirow}

\begin{document}

\title{Towards a resolved Kennicutt-Schmidt law at high redshift}

\author{J.~Freundlich\inst{1}\fnmsep\thanks{\email{jonathan.freundlich@obspm.fr}}
          \and
          F.~Combes\inst{1}
          \and
          L. J.~Tacconi\inst{2}
          \and
          M.~C.~Cooper\inst{3}
          \and
          R.~Genzel\inst{2,4,5}
          \and
          R.~Neri\inst{6}
          \and
          A.~Bolatto\inst{7}
          \and
          F.~Bournaud\inst{8}
          \and
          A.~Burkert\inst{9}
          \and
          P.~Cox\inst{6}
          \and
          M.~Davis\inst{5}
          \and
          N. M.~F\"orster Schreiber\inst{2}
          \and
          S.~Garcia-Burillo\inst{10}
          \and
          J.~Gracia-Carpio\inst{2}
          \and
          D.~Lutz\inst{2}
          \and
          T.~Naab\inst{11}
          \and
          S.~Newman\inst{5}
          \and
          A.~Sternberg\inst{12}
          \and
          B.~Weiner\inst{13}
          }

\institute{
	LERMA, Observatoire de Paris, CNRS, 61 av. de l'Observatoire, F-75014 Paris, France
	\and
	Max-Planck-Institute f\"{u}r extraterrestrische Physik (MPE), Giessenbachstrasse 1, 85748 Garching, Germany
	\and 
	Dept. of Physics \& Astronomy, Frederick Reines Hall, University of California, Irvine, CA 92697, USA
	\and
	Dept. of Physics, Le Conte Hall, University of California, CA 94720 Berkeley, USA
	\and
	Dept. of Astronomy, Campbell Hall, University of California, CA 94720 Berkeley, USA
	\and
	IRAM, 300 rue de la Piscine, F-38406 St. Martin d'Heres, Grenoble, France	
	\and
	Dept. of Astronomy, University of Maryland, College Park, MD 20742-2421, USA 
	\and
	CEA, IRFU, SAp, 91191 Gif-sur-Yvette, France
	\and
	Universit\"atsternwarte der Ludwig-Maximiliansuniversit\"at, Scheinerstrasse 1, 81679 M\"unchen, Germany,
	MPG-Fellow at MPE
	\and
	Observatorio Astron\'omico Nacional - OAN, Apartado 1143, 28800 Alcal\'a de Henares, Madrid, Spain
	\and
	Max Planck Institut f\"ur Astrophysik, Karl Schwarzshildstrasse 1, D-85748 Garching, Germany
	\and
	School of Physics and Astronomy, Tel Aviv University, Tel Aviv 69978, Israel
	\and
	Steward Observatory, 933 N. Cherry Ave., University of Arizona, Tucson, AZ 85721-0065, USA	
}

\date{Received 21 December 2012 / Accepted 24 March 2013}

\abstract
{Massive galaxies in the distant Universe form stars at much higher rates than today. Although direct resolution of the star forming regions of these galaxies is still a challenge, recent molecular gas observations at the IRAM Plateau de Bure interferometer enable us to study the star formation efficiency on subgalactic scales around redshift $z = 1.2$.
We present a method for obtaining the gas and star formation rate (SFR) surface densities of ensembles of clumps composing galaxies at this redshift, even though the corresponding scales are not resolved. This method is based on identifying these structures in position-velocity diagrams corresponding to slices within the galaxies. 
We use unique IRAM observations of the CO(3-2) rotational line and DEEP2 spectra of four massive star forming distant galaxies - EGS13003805, EGS13004291, EGS12007881, and EGS13019128 in the AEGIS terminology - to determine the gas and SFR surface densities of the identifiable ensembles of clumps that constitute them. The integrated CO line luminosity is assumed to be directly proportional to the total gas mass, and the SFR is deduced from the [OII] line. 
We identify the ensembles of clumps with the angular resolution available in both CO and [OII] spectroscopy; i.e., 1-1.5". SFR and gas surface densities 
are averaged in areas of this size, which is also the thickness of the DEEP2 slits and of the extracted IRAM slices, 
and we derive a spatially resolved Kennicutt-Schmidt (KS) relation on a scale of $\sim$ 8~kpc.
The data generally indicates an average depletion time of 1.9 Gyr, but with significant variations from point to point within the galaxies.}


\maketitle

\section{Introduction}

Ten billion years ago, between redshifts 1 and 3, observed galaxies formed their stars at rates as high as ten times that of the Milky Way today (\cite{noeske}, \cite{daddi}). This implies a more abundant gas supply, either fueled by major mergers or by a semi-continuous gas accretion (\cite{dekel2009}, \cite{dekelsari}). Recent observations of normal, massive, star-forming distant galaxies tend to show that most of these galaxies ($\geqslant 80 \%$) 
are not experiencing major mergers or interactions (\cite{tacconi2013}) and that their light profiles are similar to those of their low redshift counterparts (\cite{wuyts2011b}). Also, the Kennicutt-Schmidt (KS) relation between total molecular gas and star formation rate (SFR) surface densities is nearly linear at all redshifts (\cite{kennicutt}, \cite{wyder}, \cite{genzel}, \cite{tacconi2010}), implying that the star formation processes seem 
to be largely independent of the cosmic epoch. The average fraction of molecular gas relative to the total baryonic mass must therefore have been up to ten times higher at high redshift than in today's nearby galaxies, which is indeed corroborated by direct observations (\cite{tacconi2013}).

Numerical simulations suggest that typical massive star forming galaxies at high redshift continually accrete gas from the intergalactic medium along cold and clumpy streams stemming from the cosmic web, while the disk fragments into a few clumps
(\cite{keres}, \cite{bournaud}, \cite{dekel2009}, \cite{vandevoort}). 
Galactic disks of distant galaxies are indeed found to be fragmented in a number of giant molecular clouds, or clumps, which differs from local galaxies,  in which the molecular gas is scattered in numerous lower mass giant molecular clouds.
The high-redshift star-forming complexes 
are found to have typical scales of $\sim$1~kpc and masses up to $10^9~ \textrm{M}_\odot$, and they contribute 10-25$\%$ of the galaxy luminosity (\cite{forster}). 
Most studies only consider averaged quantities for distant galaxies, as direct observations of the star forming regions of high redshift galaxies are still challenging. 
KS relations have been derived on subgalactic scales for nearby galaxies (\cite{bigiel}, \cite{bigiel2011}, \cite{leroy2013}) but not yet extensively at high redshift. 

We use IRAM Plateau de Bure CO observations from the PHIBSS survey (\cite{tacconi2013}) and Keck DEEP2 spectra (\cite{newman}) of four massive galaxies at 
redshift $z \sim 1.2$ in order to investigate the star formation efficiency within their clumps, or groups of clumps.
More precisely, our goal is to determine the SFR and the gas mass surface densities on subgalactic scales and examine their correlation.

\begin{figure}
\centering
   \includegraphics[width=0.49\linewidth]{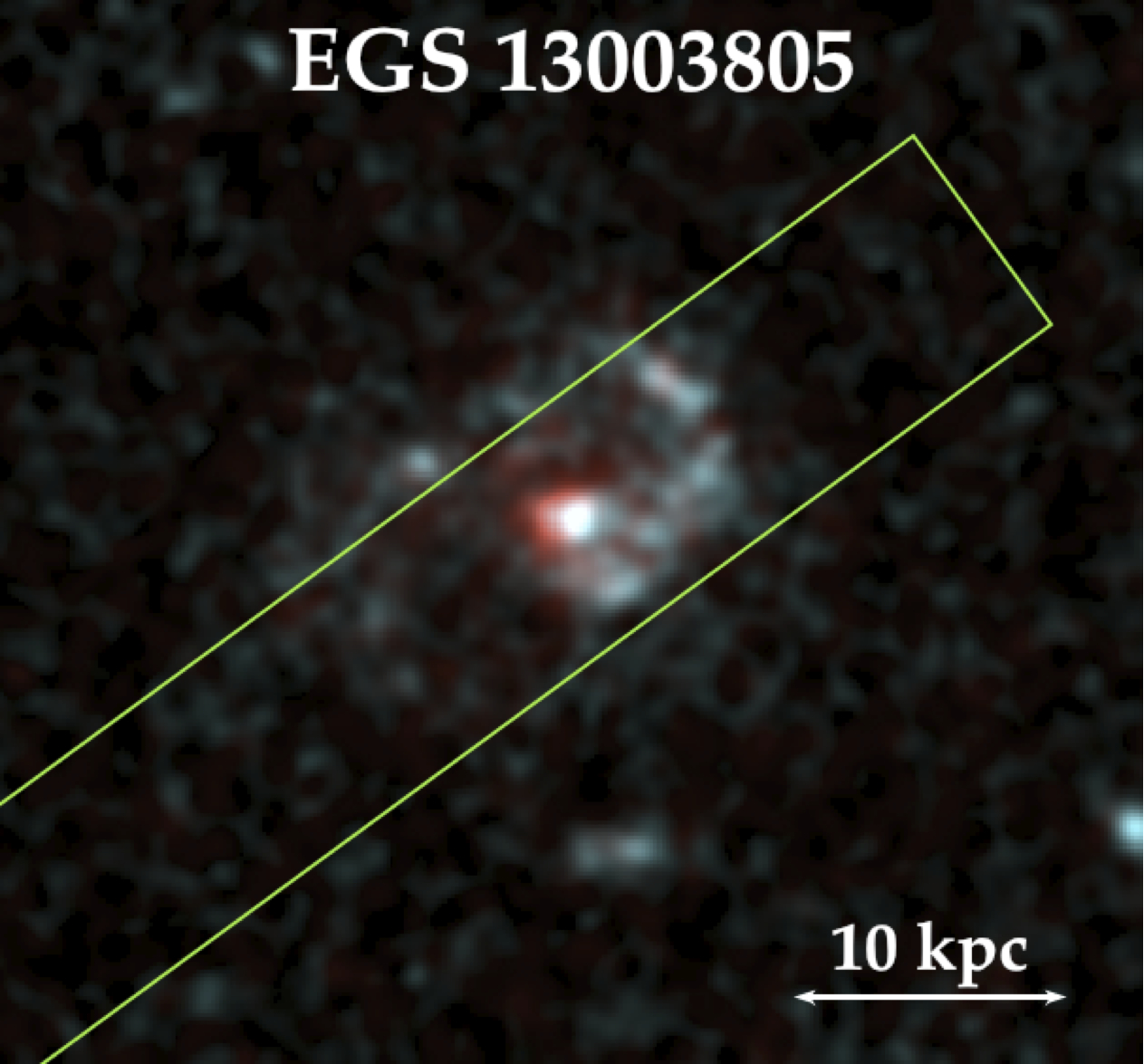}
   \includegraphics[width=0.49\linewidth]{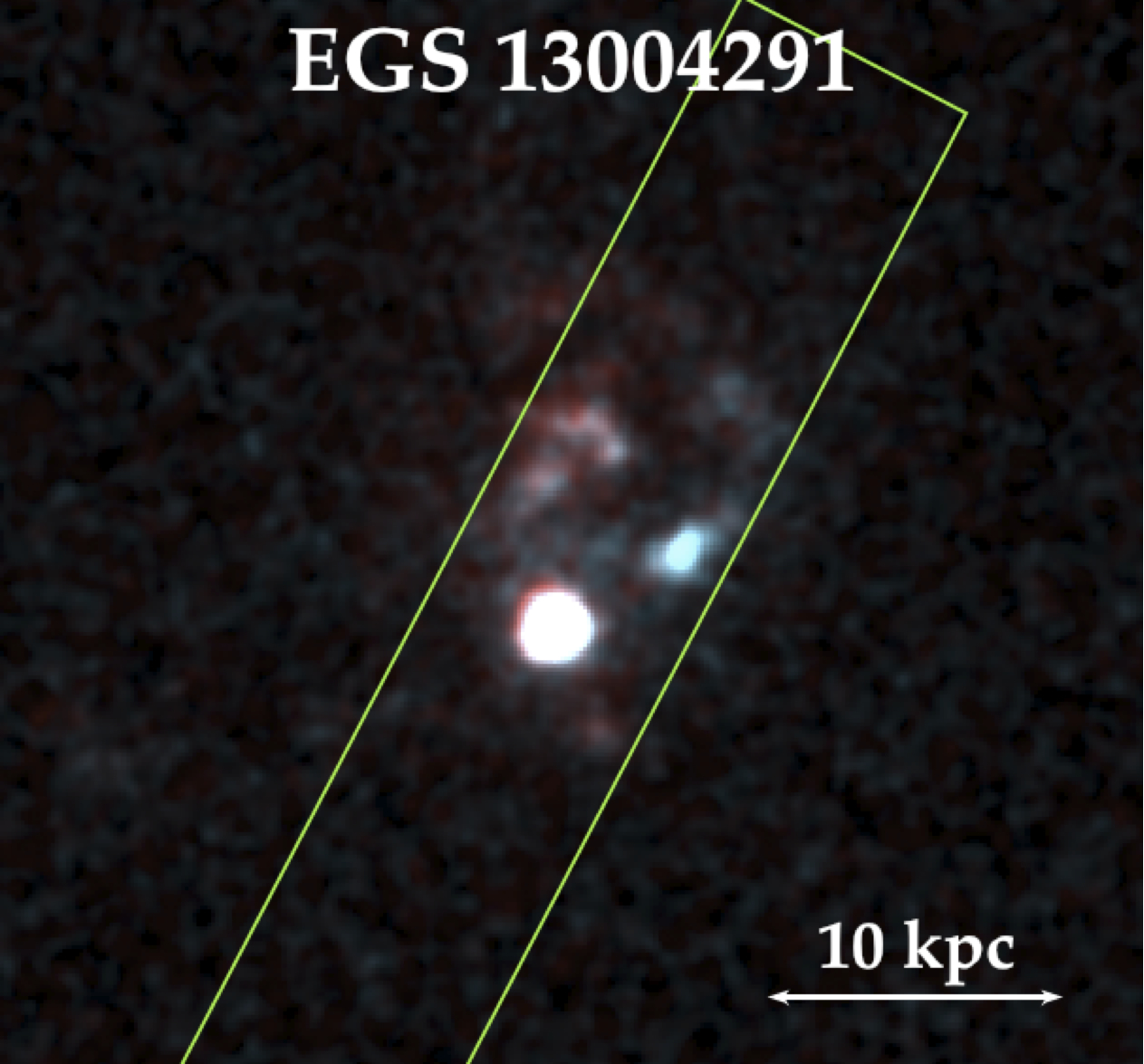}
   \includegraphics[width=0.49\linewidth]{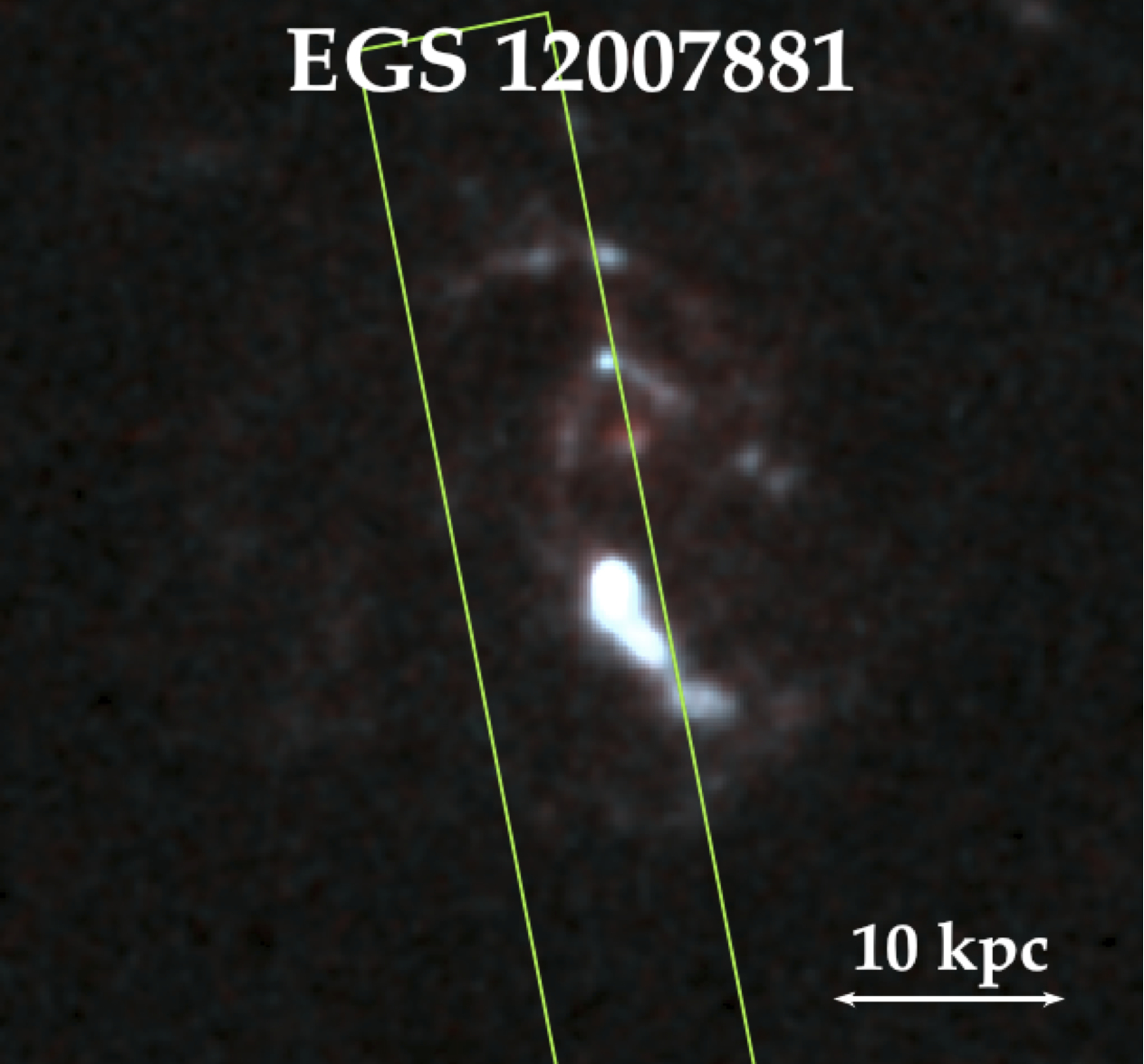}
   \includegraphics[width=0.49\linewidth]{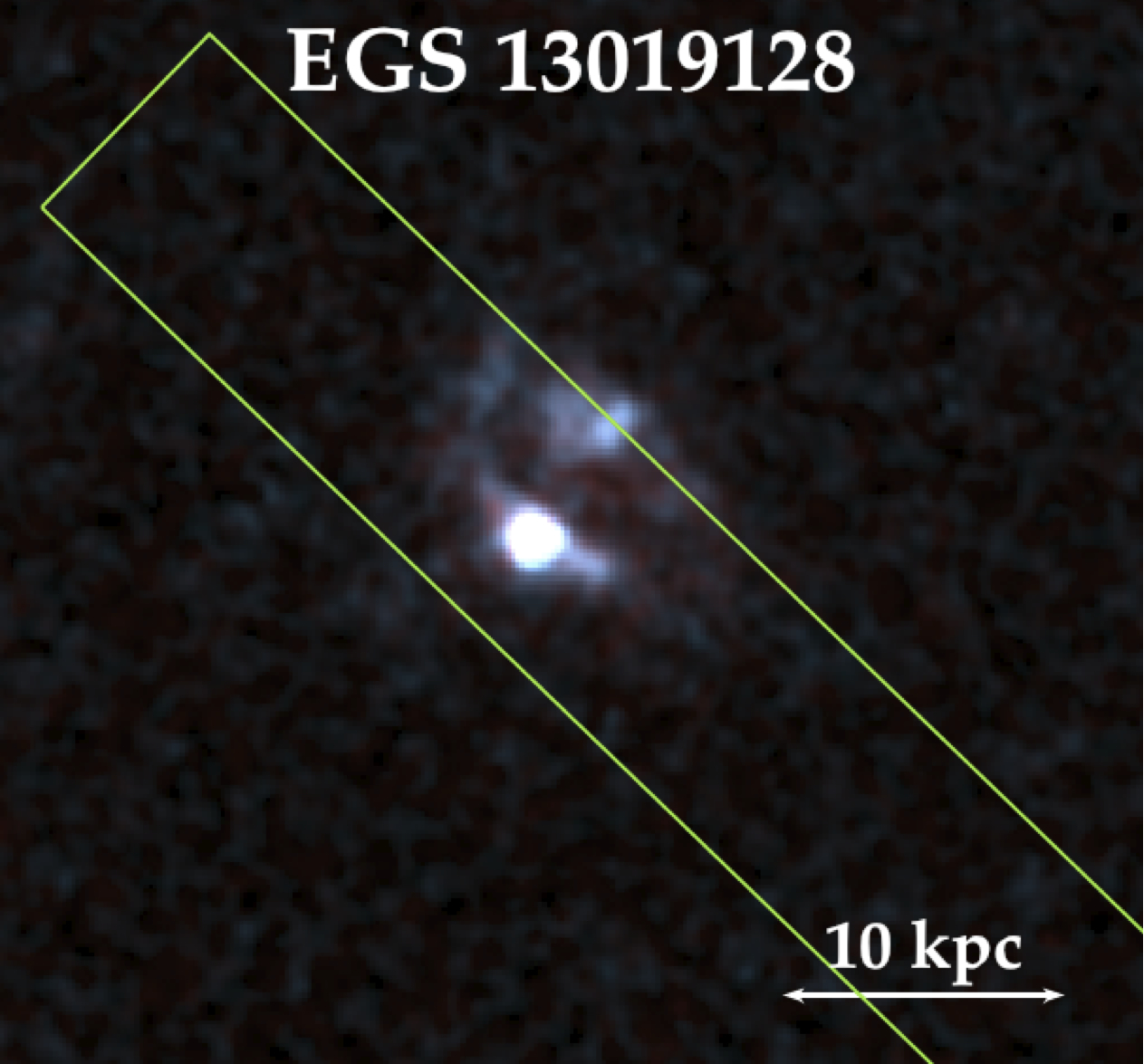}
   \caption{Composite HST images of the four galaxies studied here, combining ACS I and V bands. The green box shows the DEEP2 
slit used for each galaxy, whose size is indicated in Table \ref{table:galaxies}.
We assumed that the DEEP2 survey measured the position of the center of the galaxy more accurately than the HST, and shifted the HST image accordingly. North is up, east is left.}
   \label{galaxies}
\end{figure}

The four galaxies
were drawn from the All-wavelength Extended Groth strip International Survey (AEGIS), which provides deep imaging in all major wave bands from X-rays to radio, including Hubble Space Telescope (HST) images, and optical spectroscopy {(DEEP2/Keck)} over a large area of sky
(\cite{noeske}, \cite{davis07}, \cite{cooper2011}, 2012, \cite{newman}), thus providing a complete set of galaxies between $0.2 \leqslant z \leqslant 1.2$.
The four galaxies studied here were selected from the sample analyzed by Tacconi et al. (2010, 2013), which gathered non-major-merger,  
luminous, star forming galaxies at $z$$\sim$1.2, with stellar masses above 
$3~10^{10}~\mathrm{M}_{\odot}$ and SFR above $40~\mathrm{M}_{\odot}~\mathrm{yr}^{-1}$. These four galaxies - named EGS13003805, 
EGS13004291, EGS12007881, and EGS13019128 in the AEGIS terminology - correspond to 
the massive end of the `normal' star forming galaxies at $z$$\sim$1.2 (\cite{tacconi2010}). 
Figure \ref{galaxies} shows them in the I and V bands of the HST Advanced Camera for Surveys (ACS), as well as the DEEP2 slits  we used. 
To compute the physical distances, we adopt a cosmology with $\Omega_\Lambda = 0.73$, $\Omega_m = 0.27$, and $\mathrm{H}_0 = 70 ~\mathrm{km~s}^{-1}\mathrm{Mpc}^{-1}$.

\section{Determination of the total gas mass and the SFR}

\subsection{The gas mass}

We use high resolution observations of the CO(3-2) transition shifted into the 2 mm band for $z$$\sim$1.2 sources carried at the IRAM Plateau de Bure, including the most extended 
`A' configuration. The angular resolution obtained is between 0.5" and 1.5" for all galaxies (Table \ref{table:galaxies}), corresponding to physical scales between 4 and 13 kpc. 
The CO luminosity associated to any region of flux 
can be expressed as 
\[
\left(\frac{\mathrm{L}^\prime_{\mathrm{CO}}}{\mathrm{K.km.s}^{-1}\mathrm{.pc}^2}\right) = \frac{3.25 \timesÊ10^7}{\left(1+z\right)^{3}}~\left(\frac{S_{\mathrm{CO}} \Delta V}{\mathrm{Jy.km.s}^{-1}}\right)~ \left(\frac{\nu_{obs}}{\mathrm{GHz}}\right)^{-2}~ \left(\frac{\mathrm{D}_\mathrm{L}}{\mathrm{Mpc}}\right)^2 
\]
where $S_{CO} \Delta V$ is the  velocity integrated flux, $\nu_{obs}$ the observed frequency, and $D_L$ the luminosity distance (\cite{solomon97}). 
To derive H$_2$ masses, we follow the approach used by \cite{tacconi2010}. This assumes
a Milky-Way-like conversion factor of $\alpha = 3.2~\mathrm{M}_{\odot}~ (\mathrm{K~km~s}^{-1}~\mathrm{pc}^2)^{-1}$ 
between the CO(1-0) luminosity and the H$_2$ mass, a factor 1.36 to account for interstellar helium, 
and a further correction by a factor 2 for the CO(3-2)/CO(1-0) luminosity ratio. 
The total gas masses of the four studied galaxies are indicated in Table \ref{table:galaxies}.

\subsection{The star formation rate}
\label{section:sfr}

The H$\alpha$ recombination line is the most direct and reliable tracer of young massive stars, and thus of the SFR (\cite{kennicutt98}). Since this line lies in the middle of a low atmospheric transmission band for the four galaxies considered here, ground-based H$\alpha$ spectroscopy is impossible, and we estimate the SFR from the collisionally excited [OII] line (rest frame wavelength $3727~ \AA$, shifted around $8200~ \AA$ for $z$$\sim$1.2 sources).

Though the [OII] forbidden line luminosity is not directly coupled to the ionizing luminosity, it is possible to establish it empirically as a quantitative SFR tracer.
The Kennicutt (1998b) [OII] SFR calibration has been broadly used, but does not take reddening and metallicity into account. 
Here we use the calibration proposed by Kewley et al. (2004), which includes these two effects. The four galaxies studied here are particularly massive and should thus belong to the observed metallicity plateau at 12+log(O/H) = 9.07 (\cite{mannucci}). Assuming this metallicity for all galaxies, the Kewley et al. (2004) [OII] SFR calibration gives
\[
\left(\frac{\mathrm{SFR}}{\mathrm{M}_\odot .\mathrm{yr}^{-1}}\right) \simeq 3.5 \times 10^{-8}~\left(\frac{\mathrm{L}_{\mathrm{[OII]}}}{\mathrm{L}_\odot}\right)
\]
where L$_{\mathrm{[OII]}}$ is the intrinsic [OII] line luminosity. This calibration is expected to follow the reliable standard \cite{kennicutt98} H$\alpha$ SFR calibration with a scatter of 0.03-0.05 dex (\cite{kewley}).

To determine L$_{\mathrm{[OII]}}$, we use spectra provided by the Keck DEEP2 survey, obtained with the high spectral resolution DEIMOS spectrograph (\cite{faber}) and taken along slits with a typical size of $5'' \times 1''$. The data was reduced using a dedicated pipeline (\cite{newman}, \cite{cooper}),
and a velocity resolution of about $50~ \mathrm{km~ s}^{-1}$ ($R = \lambda/\Delta \lambda \approx 5000$) enables resolving the  [OII] line (\cite{davis03}).
DEEP2 spectra are not flux-calibrated, and 
we use CFHT I band magnitudes (CFH12K camera) to calibrate the galaxy-integrated ``1D" spectrum (\cite{coil}).

\begin{table*}
\caption{Some global properties and observational details for the four galaxies studied here: (1) the redshift, as determined in the AEGIS catalogues from the [OII] line using DEEP2 spectra; (2) the angular distance $D_A$ and the luminosity distance $D_L$; (3) the SFR from the [OII] line luminosity in the galaxy-integrated ``1D'' spectrum, obtained according to Kewley et al. (2004) calibration; (4) the extinction-corrected SFR obtained from 24 $\mu$m and UV continuum by Tacconi et al. (2013); (5) the SFR extrapolated from the empirical calibration of Moustakas et al. (2006);  
(6) the gas mass derived from the IRAM Plateau de Bure observations; (7) the total star mass obtained by SED fittings from CFHT B, R, and I bands photometric data using \texttt{kcorrect} (\cite{blanton}), corrected for a Universe with $\mathrm{H}_0 = 70 ~\mathrm{km.s}^{-1}.\mathrm{Mpc}^{-1}$; (8) the gas fraction $\mathrm{f}_{\mathrm{gas}} = \mathrm{M}_{\mathrm{gas}}/(\mathrm{M}_{\mathrm{gas}}+\mathrm{M}_{\mathrm{star}})$; (9) the optical half light radius $R_{1/2}$ as indicated in Tacconi et al. (2013);
(10) the surface densities associated to the gas mass, the star formation rate, and the star mass, calculated within $R_{1/2}$, for example as $\Sigma_{\mathrm{gas}} = 0.5~ \mathrm{M}_{\mathrm{gas}}/\pi R_{1/2}^2$;  (11) H$\alpha$, visible, and [OII] extinctions as derived from the \texttt{kcorrect} SED reconstruction ; and (12) the extinction ratio  $A_{H\alpha}/A_{[OII]}$. $A_V $, $A_{[OII]} $,  $A_{H\alpha}/A_{[OII]}$, SFR$_{24\mu m + \mathrm{UV}}$ and SFR$_{\mathrm{[OII],B}}$ are given for information only. O'Donnell (1994) predicts that $A_{H\alpha}/A_{[OII]}$ is around 1.86 for a diffuse interstellar medium, and the variations here observed from one galaxy to the other are representative of the quality of the \texttt{kcorrect} fit. The comparison between  SFR$_{\mathrm{[OII]}}$, SFR$_{24\mu m + \mathrm{UV}}$, and SFR$_{\mathrm{[OII],B}}$ gives an idea of the uncertainties in the SFR.
}
\label{table:galaxies}
\centering
\begin{tabular}{lllll}
\hline 
\hline
Properties &  EGS  13003805 & EGS 13004291  & EGS 12007881 & EGS 13019128\\
\hline
$z$ $^{(1)}$										&$1.2318 $		&$1.19705$		&$1.16105$		&$1.3494$\\
$D_A$	 [Mpc] $^{(2)}$								&$ 1753.0$		&$1745.3$		&$1736.4$		&$1772.8$\\
$D_L$	 [Mpc] $^{(2)}$								&$8731.5 $		&$8424.7$		&$8108.8$		&$9785.1$\\
Scale [kpc/"]										&$8.50$			&$8.46$			&$8.42$			&$8.60$	\\		
\hline
DEEP2 slit  size &$6.863^{\prime\prime}\times1^{\prime\prime}$&$5.149^{\prime\prime}\times1^{\prime\prime}$& $5.762^{\prime\prime}\times1^{\prime\prime}$&$10.343^{\prime\prime}\times1^{\prime\prime}$\\
DEEP2 slit orientation (PA) & $-53.858^{\circ}$ & $-27.649^{\circ}$ &$11.142^{\circ}$ &$46.142^{\circ}$\\
IRAM CO(3-2) beam & $0.8^{\prime\prime}\times 0.6^{\prime\prime}$&$0.6^{\prime\prime}\times0.5^{\prime\prime}$&$1.1^{\prime\prime}\times 0.9^{\prime\prime}$ & $1.6^{\prime\prime}\times 1.4^{\prime\prime}$\\
\hline
SFR$_{\mathrm{[OII]}}$  [$\mathrm{M}_\odot~\mathrm{yr}^{-1}$] $^{(3)}$		&$  98$			&$182$			&$119$			&$202$\\
SFR$_{\mathrm{UV}+24\mu m}$ [$\mathrm{M}_\odot~\mathrm{yr}^{-1}$] $^{(4)}$	& 200	& 630	& 94				& 87		\\
SFR$_{\mathrm{[OII], B}}$ [$\mathrm{M}_\odot~\mathrm{yr}^{-1}$] $^{(5)}$		& 63 				& 231			& 265			& 213	\\
\hline
$\mathrm{M}_{\mathrm{gas}} $ [$10^{11}~\mathrm{M}_\odot$] $^{(6)}$	&$2.2 $	&$ 2.8$			&$1.3$			&$1.2$\\
$\mathrm{M}_{\mathrm{star}} $ [$10^{11}~\mathrm{M}_\odot$] $^{(7)}$	&$3.4 $	&$5.0$			&$5.0$			&$3.8$\\
$\mathrm{f}_{\mathrm{gas}}$ $^{(8)}$							& $ 0.39$ & $0.36$	&$0.21$			&$0.24$\\
\hline
$R_{1/2} $ [$\mathrm{kpc}$] $^{(9)}$					&$  5.7$			&$3.1$			&$5.7$			&$5.2$\\
$\mathrm{log}_{10}(\Sigma_{\mathrm{gas}}/[\mathrm{M}_\odot~\mathrm{pc}^{-2}]) $ $^{(10)}$&$  3.03$	&$3.67$			&$2.80$			&$2.86$\\
$\mathrm{log}_{10}(\Sigma_{\mathrm{SFR}}/[\mathrm{M}_\odot~\mathrm{yr}^{-1}~\mathrm{kpc}^{-2}]) $ $^{(10)}$&$  -0.193$&$0.604$&$-0.095$	&$0.204$\\
$\mathrm{log}_{10}(\Sigma_{\mathrm{star}}/[\mathrm{M}_\odot~\mathrm{pc}^{-2}]) $ $^{(10)}$&$  3.22$	&$3.92$			&$3.39$			&$3.36$\\
\hline
$A_{H\alpha} $ $^{(11)}$								&$  1.25$			&$1.28$ 			&$1.06$			&$1.21$\\
$A_V $ $^{(11)}$									&$  1.56$			&$1.60$			&$1.30$			&$1.50$\\
$A_{[OII]} $ $^{(11)}$									&$  2.28$			& $2.26$			&$2.04$			&$2.19$\\
$A_{[OII]} / A_{H\alpha}$ $^{(12)}$						&1.82			&1.77			&1.92			&1.81\\
\hline
\end{tabular}
\end{table*}

\begin{table*}
\caption{Some properties of the clumps of EGS 13004291, EGS 13003805, EGS 12007881, and EGS 13019128 obtained from measurements in the position-velocity diagrams: (1) the gas mass obtained as a fraction of the total mass of the galaxy from the CO position-velocity diagram; (2) the SFR obtained similarly from the [OII] position-velocity diagram, normalized with the total SFR determined according to section \ref{section:sfr}; (3) the derived gas mass and SFR surface densities averaged over the same area of 1'' in diameter; (4) the depletion time $\mathrm{t}_{\mathrm{depl}}=\mathrm{M}_{\mathrm{gas}}/\mathrm{SFR}$; and (5) the FWHM obtained from the IRAM position-velocity diagram with a Gaussian fit in the direction of the slice (for information only). The FWHM roughly corresponds to the IRAM beam size, so we preferred to use the same 1" size for all clumps in the different calculations. Using the FWHM instead of a constant 1" diameter to calculate the surface densities gives a much larger scatter, but actually does not change the constant depletion time fit and the values of t$_{\mathrm{depl}}$. The clumps are numbered from bottom to top according to the horizontal separation lines of Figure \ref{pvdiag}.}
\label{table:clumps}
\centering
\begin{tabular}{lcccccc}
\hline
\hline
Clump  &  $ \mathrm{M}_{\mathrm{gas}}$ [$10^{10}$ M$_\odot$] $^{(1)}$ & SFR [M$_\odot$~yr$^{-1}$] $^{(2)}$& $ \mathrm{log}_{10}\left(\frac{\Sigma_{\mathrm{gas}}}{\mathrm{M}_\odot~\mathrm{pc}^{-2}}\right)$ $^{(3)}$& $ \mathrm{log}_{10}\left(\frac{\Sigma_{\mathrm{SFR}}}{\mathrm{M}_\odot~\mathrm{yr}^{-1}~\mathrm{kpc}^{-2}}\right)$ $^{(3)}$ & $\mathrm{t}_{\mathrm{depl}}$ [Gyr] $^{(4)}$ & FWHM [$^{\prime\prime}$] $^{(5)}$\\
\hline
 EGS 13003805-I 		& $11.85$ 	& $38.23$  & $3.019$ & $-0.473$ & 3.10 & $0.93$\\
 EGS 13003805-II 		& $6.92$ 		& $34.75$  & $2.785$& $-0.514$ & 1.99 & $0.78$\\
 EGS 13003805-III 		& $3.18$ 		& $24.99$  & $2.447$& $-0.657$ & 1.27 & $0.63$\\
\hline
  EGS 13004291-I 		& $2.37$ 		& $11.37$  & $2.324$ & $-0.995$ & 2.09 & $0.78$\\
 EGS 13004291-II 		& $5.80$ 		& $27.08$  & $2.713$ & $-0.618$ & 2.14& $0.64$\\
 EGS 13004291-III 	 	& $7.75$ 		& $26.16$  & $2.838$ & $-0.633$ & 2.96 & $0.74$\\
 EGS 13004291-IV 	 	& $4.96$ 		& $23.92$  & $2.644$ & $-0.672$ & 2.07 & $0.79$\\
 EGS 13004291-V 		& $6.28$ 		& $33.83$  & $2.761$ & $ -0.522$ & 1.92 & $0.61$\\
 EGS 13004291-VI 	 	& $3.99$ 		& $23.45$  & $2.550$ & $-0.681$    & 1.70 & $1.1$ \\
 \hline
 EGS 12007881-I 		& $5.33$ 	& $25.38$  & $2.680$ & $-0.642$ & 2.10 & $1.35$\\
 EGS 12007881-II 		& $7.26$ 	& $41.88$  & $2.814$& $-0.425$ & 1.73 & $1.36$\\
 EGS 12007881-III 		& $1.41$ 	& $51.05$  & $2.10$& $-0.339$ & 0.28 & $1.57$\\
\hline
 EGS 13019128-I 		& $1.70$ 	& $4.03$ & $2.165$ & $-1.460$ & 4.22 & $0.83$\\
 EGS 13019128-II 		& $0.58$ 	& $57.23$ & $1.697$& $-0.307$ & 0.10 & $1.14$\\
 EGS 13019128-III 		& $7.08$ 	& $52.78$ & $2.785$& $-0.342$ & 1.34 & $1.67$ \\
 EGS 13019128-IV		& $5.25$ 	& $47.51$ & $2.655$& $-0.388$ & 1.10 & $1.60$ \\
 EGS 13019128-V		& $0.78$ 	& $32.32$ & $1.828$& $-0.555$ & 0.24 & $1.62$\\
  EGS 13019128-VI		& $2.60$ 	& $5.89$ & $2.351$& $-1.295$ & 4.42 & $1.86$\\
\hline
\end{tabular}
\end{table*}

\begin{figure*}
\centering
   \includegraphics[width=0.49\linewidth]{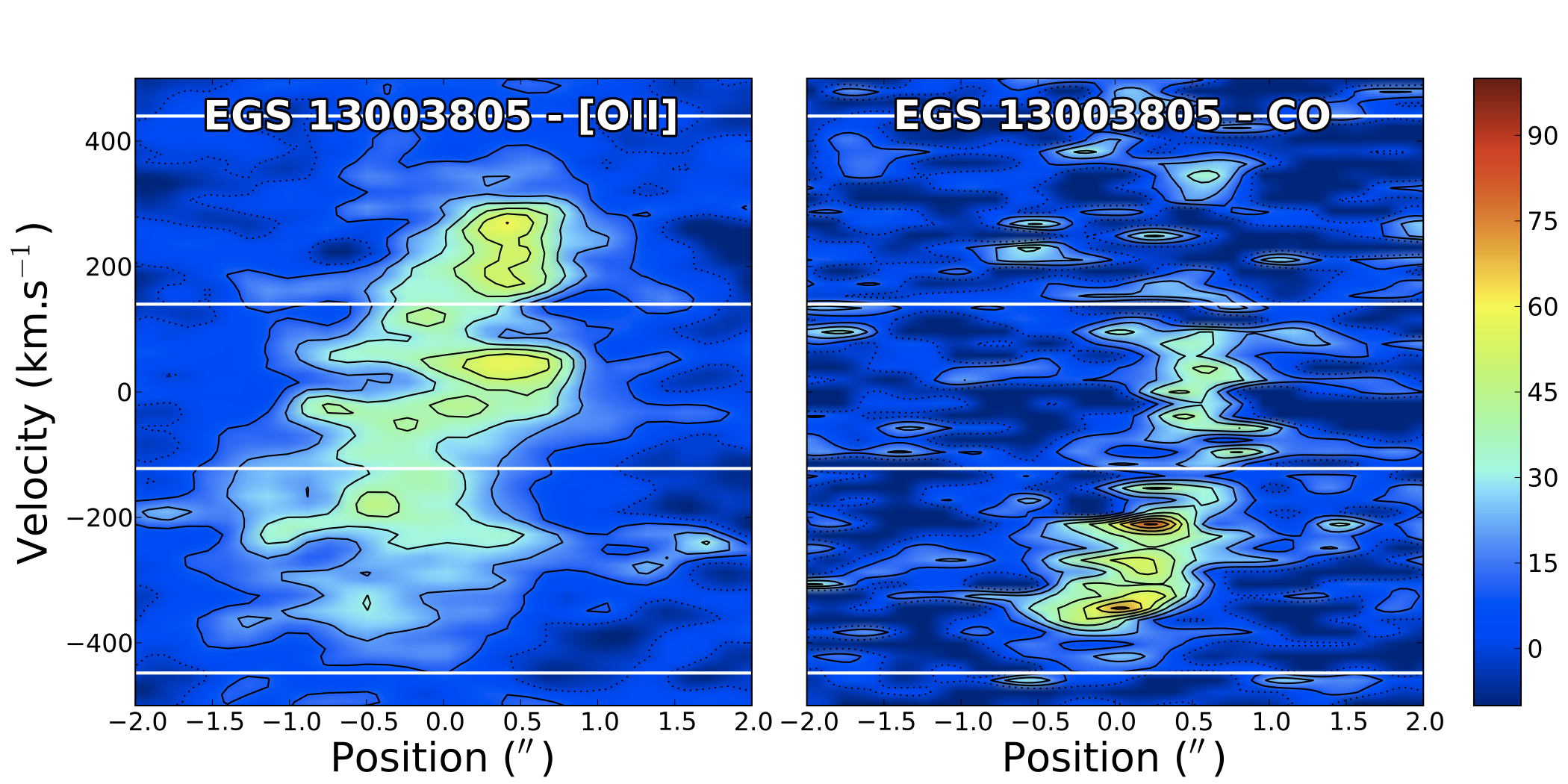}
   \includegraphics[width=0.49\linewidth]{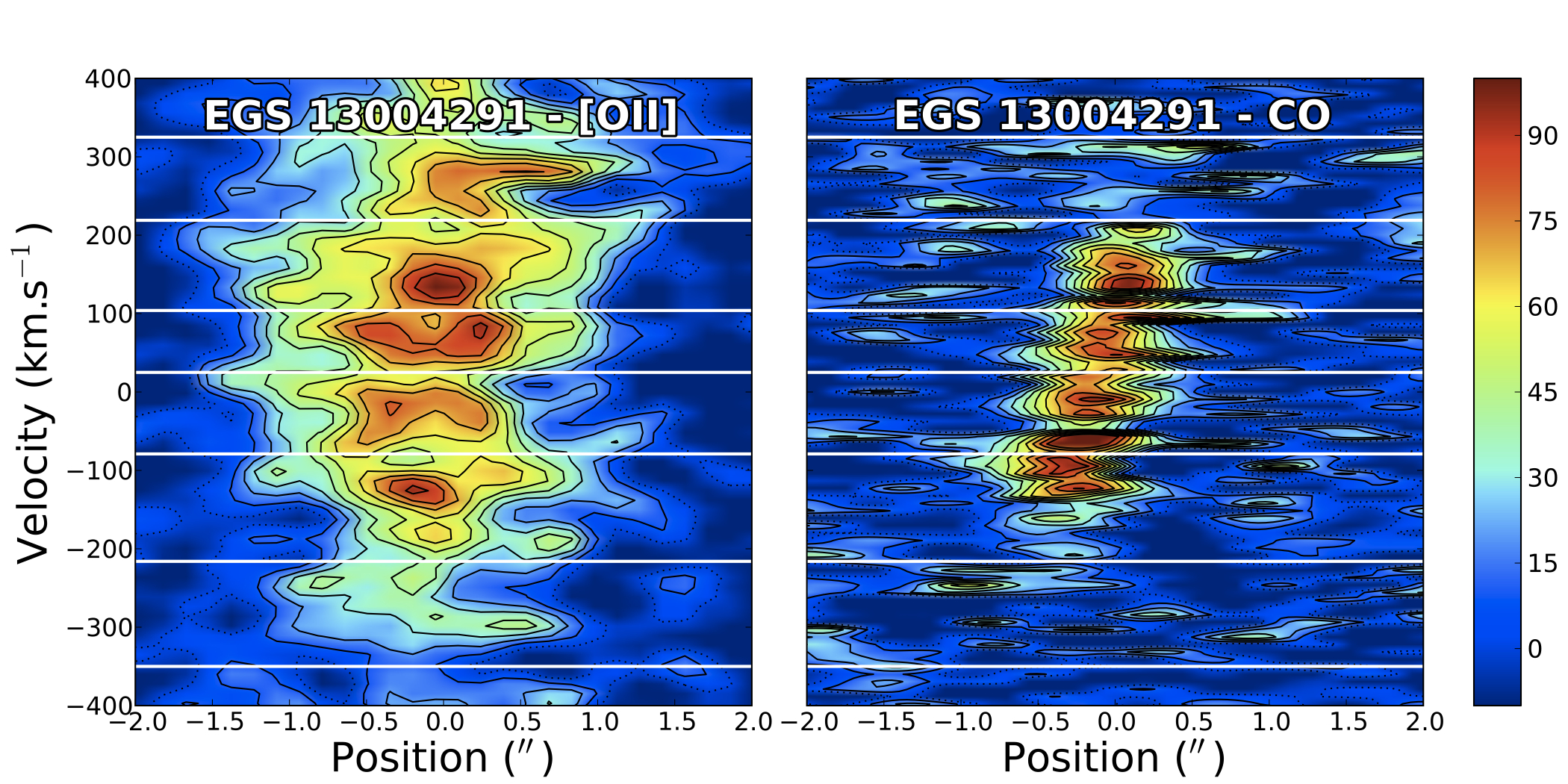}
   \includegraphics[width=0.49\linewidth]{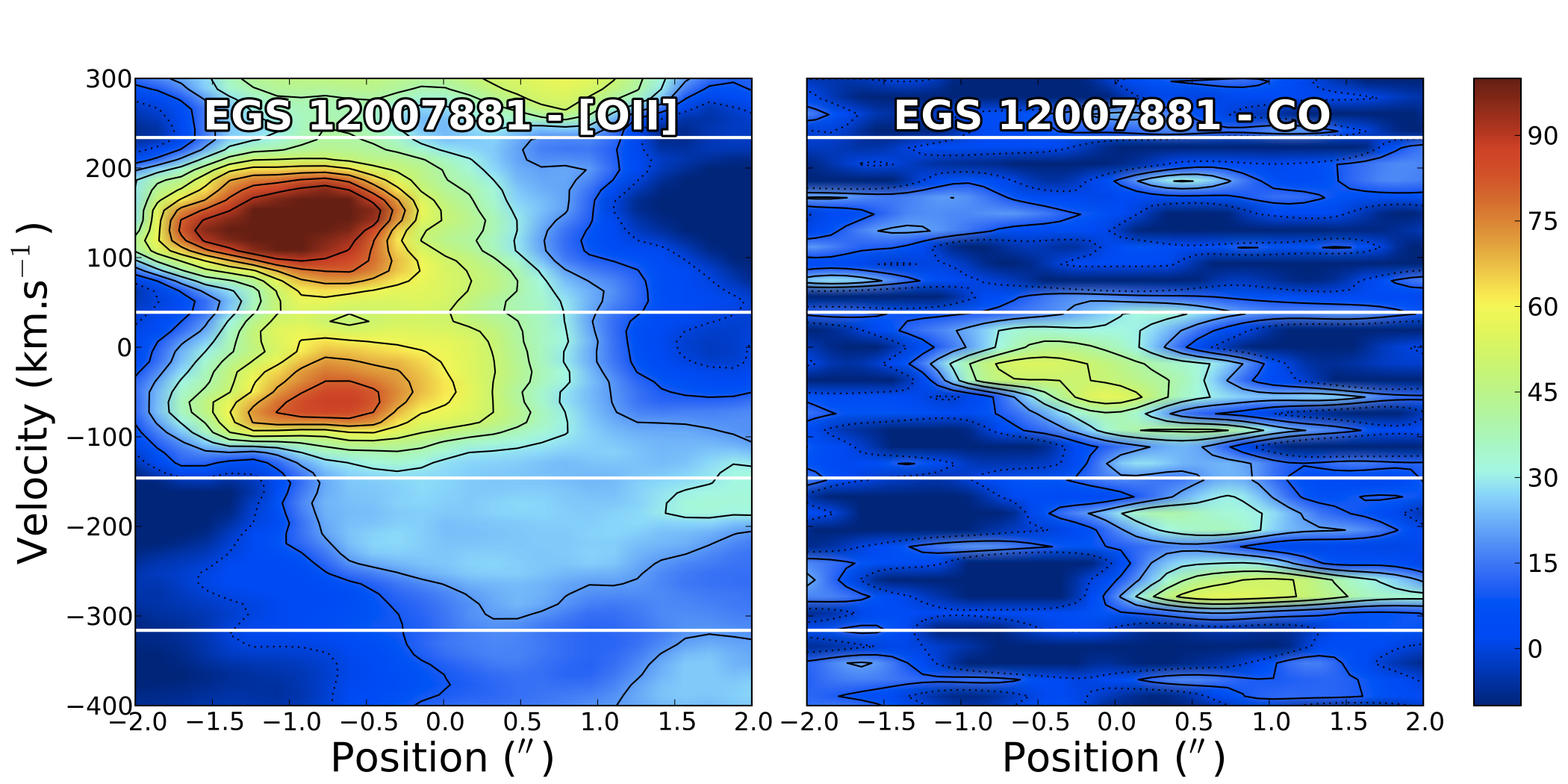}
   \includegraphics[width=0.49\linewidth]{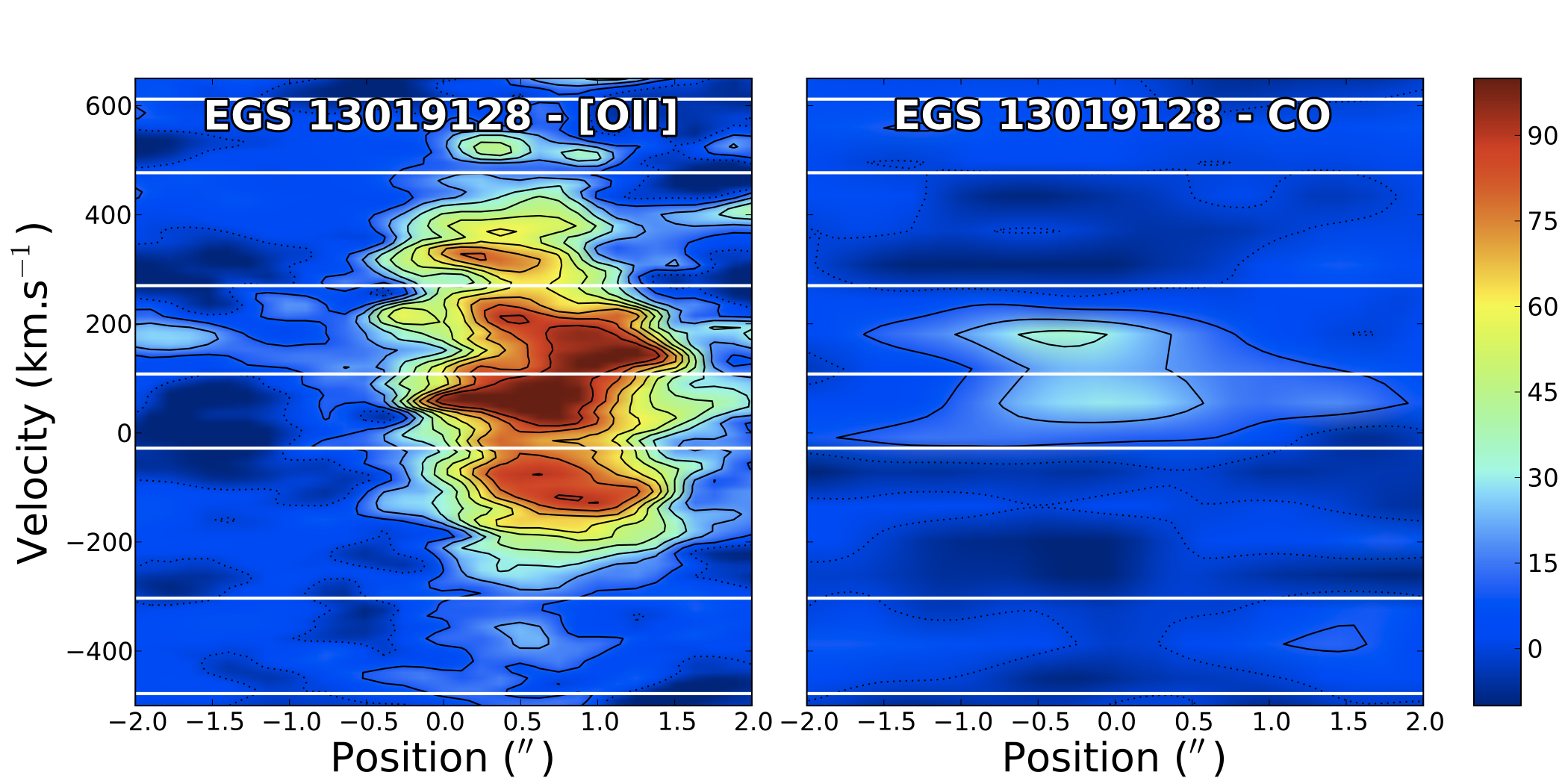}
   \caption{[OII] line and CO luminosities (respectively left and right panels for each galaxy) in position-velocity planes corresponding to the DEEP2 slits. Smoothed ensembles of clumps are separated by eye along the vertical axis, as shown with the white horizontal lines. The [OII] diagrams were normalized in order to have fluxes proportional to the SFR, but all galaxies share the same arbitrary color scales.
One arcsecond corresponds approximately to 8.5 kpc (Table \ref{table:galaxies}).}
\label{pvdiag}%
\end{figure*}

The value of L$_{\mathrm{[OII]}}$ has to be corrected for dust extinction. The Galactic extinction is taken into account in the CFHT magnitudes, since they already include the dust corrections of Schlegel, Finkbeiner \& Davis (1998), but the dust present in the galaxies themselves is an even greater cause of extinction. 
The dust distribution was derived from the SED of the galaxies, obtained from the B, R, and I band CFHT magnitudes 
with the \texttt{kcorrect} software developed by \cite{blanton} (version \texttt{v4\_2}). This software finds the best non-negative linear combination of a number of template star formation histories fitting the CFHT magnitudes. The templates are based on stellar population synthesis models, which enables the software to provide approximate physical quantities related to the star formation history. It notably provides SED curves with and without 
extinction, from which we can deduce the extinction A$_{[OII]}$ at the [OII] wavelength. 
The values for A$_{[OII]}$ are indicated in Table \ref{table:galaxies}, along with the visible extinction $A_V$ and the extinction $A_{H\alpha}$ at H$\alpha$ wavelength, obtained from the SED curves. The $A_{H\alpha}/A_{[OII]}$ ratio varies up to 10\% from galaxy to galaxy but remains comparable to the value predicted by O'Donnell (1994) for a diffuse interstellar medium: $A_{H\alpha}/A_{[OII]} =1.86$. Since this ratio should remain constant for the different galaxies, the variations give an idea of the quality of the \texttt{kcorrect} fit.
The obtained extinctions are in approximate agreement with observations (e.g. \cite{forster}), and the SFRs deduced from the [OII] luminosity are given in Table \ref{table:galaxies} as SFR$_{\mathrm{[OII]}}$.

Moustakas et al. (2006) present an alternative empirical [OII] SFR calibration parametrized in terms of the B-band luminosity. This calibration is intended to remove, on average, the systematic effects of reddening and metallicity, as well as to reduce the scatter in the resulting SFR values. The SFR of the four galaxies studied here were also determined from UV and IR luminosities by Tacconi et al. (2013). Because UV light directly traces unobscured star formation and IR 24 $\mu$m emission originates in small dust grains mainly heated by UV photons from young stars, the combination of the two is a sensitive tracer of the global SFR (\cite{leroy}, 2012). 
Tacconi et al. (2013) derive the global SFR of the four galaxies studied here from a combination of UV and Spitzer 24 $\mu$m luminosities with the methods of Wuyts et al. (2011b). 
The SFR obtained through these two calibrations are given for information in Table~\ref{table:galaxies}. The values  differ up to 3$\sigma$ between the three methods, which gives an idea of the uncertainty of the SFR measurements. Since it gives a lower standard deviation, we use the Kewley et al. (2004) calibration in our study. However, [OII] measurements may not reveal some dust-embedded star forming regions, which probably explains the higher values of the SFR traced by UV and IR luminosities determined by Tacconi et al. (2013).

\section{A resolved Kennicutt-Schmidt law}

HST images of Figure \ref{galaxies} reveal kpc-sized clumps within diffuse regions, but these clumpy features ($\sim$ 0.1") are  
smoothed out at DEEP2 and IRAM resolutions, whose ranges are respectively 0.6-1.0" and 0.5-1.6".  
Spectroscopy, however, helps separate different components, thanks to their kinematics.
DEEP2 spectra correspond to position-velocity diagrams (PV diagrams) along the galaxy major axis (\cite{davis07}).
Figure \ref{pvdiag} compares the [OII] position-velocity diagrams with the corresponding slices in CO(3-2), both with the same
1" width. 
Smoothed 1"-sized ensembles of clumps are separated by eye along the velocity axis of the PV diagrams. The characteristics of these ensembles are given in Table \ref{table:clumps}, and we aim at describing the star formation efficiency within them.
We tried to compensate for the substructure separation by eye by taking all identifiable ensembles of clumps into account.

\begin{figure}
\centering
   \includegraphics[width=0.6\linewidth]{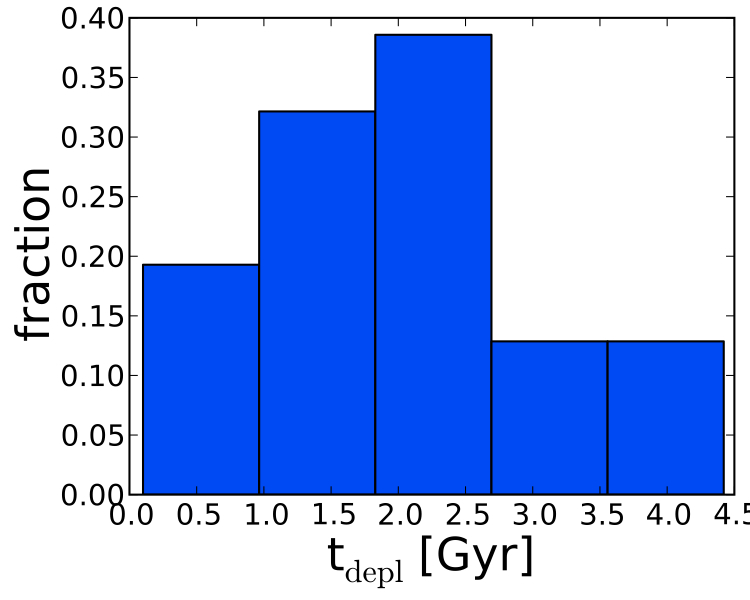}
   \caption{Distribution of the depletion time for the 16 regions identified in the four galaxies studied here, using the Kewley et al. (2004) [OII] SFR calibration. The distribution has a mean of $\mathrm{t}_{\mathrm{depl}} = 1.9$~Gyr and a standard deviation of 1.2~Gyr, for a median value of 2.0~Gyr. Since we identified 18 ensembles of clumps, the standard error of the mean is 0.3~Gyr.}
   \label{tdepl}
\end{figure}

\begin{figure}
\centering
   \includegraphics[width=1\linewidth]{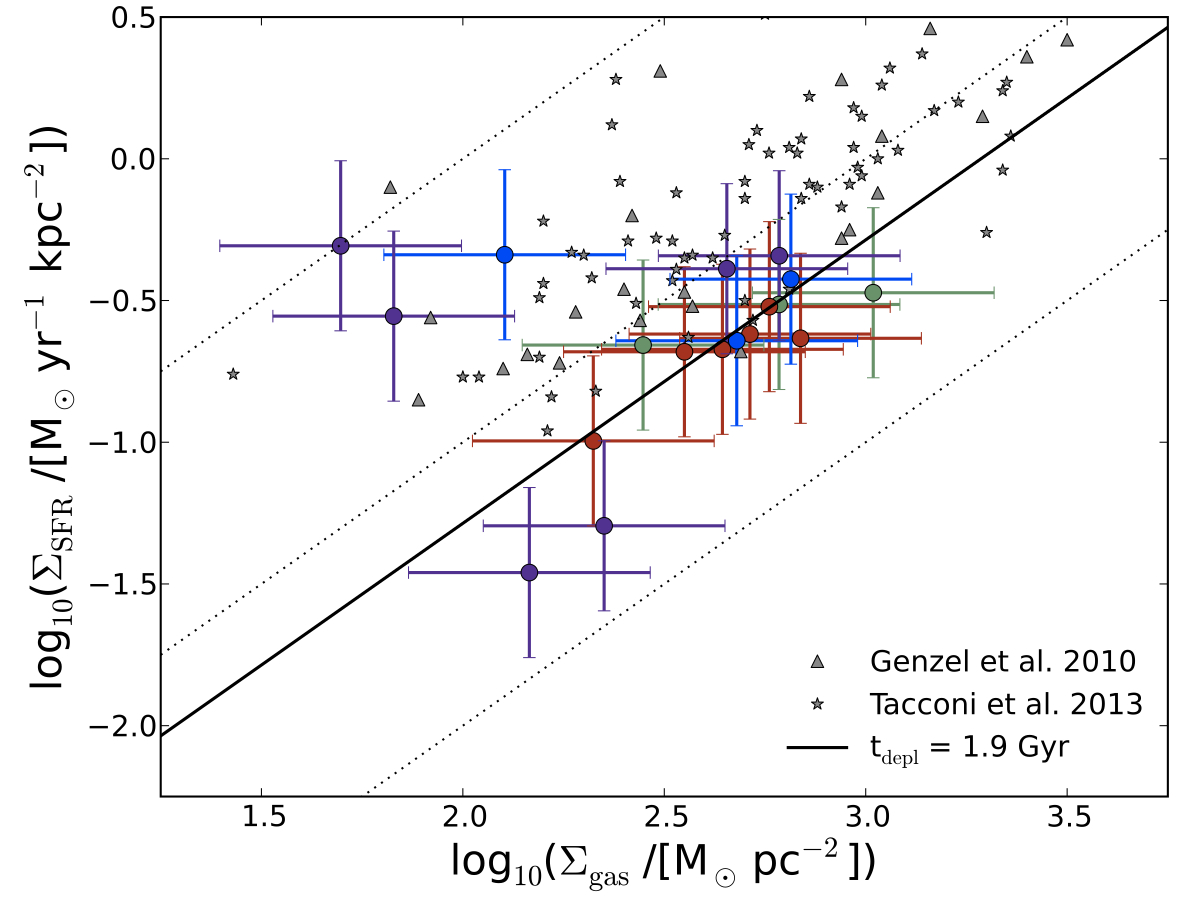}
   \caption{Kennicutt-Schmidt diagram for 1" ensembles of clumps of EGS13003805 (green), EGS13004291 (red), EGS12007881 (blue), and EGS13019128 (purple), using the Kewley et al. (2004) [OII] SFR calibration. The dotted diagonal lines correspond to constant gas depletion times of 0.1, 1, and 10 Gyr from top to bottom, and the solid black line to a constant depletion time equal to the mean depletion time of the clumps, t$_{\mathrm{depl}}$=1.9 Gyr. 
The error bars of $0.3$ dex yield a reduced $\chi^2$ close to $1$ and correspond to a factor 2 uncertainty, which is a lower estimate. The gray data points from Genzel et al. (2010) and Tacconi et al. (2013) are indicated for comparison with whole galaxies.
}   
\label{kentot}%
\end{figure}

From the CO and [OII] lines, we estimate the gas mass and the SFR contained in areas of 1" in diameter (corresponding to the width of the slice and approximately to the size of the ensembles of clumps), and obtain the corresponding averaged surface densities, $\Sigma_{\mathrm{gas}}$ and $\Sigma_{{\mathrm{SFR}}}$. 
As shown in Figure \ref{tdepl}, the depletion time ($\mathrm{t}_{\mathrm{depl}}=\mathrm{M}_{\mathrm{gas}}/\mathrm{{SFR}}$) is equal to 1.9$\pm$0.3~Gyr, to be compared with results for samples of whole galaxies (2.1 Gyr at z = 0 in \cite{kennicutt98}, 0.5-1.5 Gyr from z$\sim$2 to z$\sim$0 in \cite{genzel}, 1.05$\pm$0.74 Gyr in \cite{saintonge2011}, $\sim$0.7 Gyr at z=1-3 in \cite{tacconi2013}) and samples of subgalactic regions at low redshift (2.0$\pm$0.8 Gyr in \cite{bigiel} and 2.35 Gyr in \cite{bigiel2011}).
The resulting KS diagram
is displayed in Figure \ref{kentot}. 
The data points scatter around the line of constant depletion time equal to $1.9$ Gyr (such a line corresponding to a power law $\Sigma_{\mathrm{SFR}}\propto\Sigma_{\mathrm{gas}}^N$ of exponent N=1). 
However, the depletion time is locally very different from one data point to the other, suggesting that the star formation scaling laws are different from one ensemble of clumps to the next within a galaxy.  
The scatter is comparable to the $\sim$0.3 dex scatter observed for resolved local galaxies (\cite{bigiel}, 2011).
The corresponding values obtained with the SFR from Tacconi et al. (2013) and through the B-band calibration of Moustakas et al. (2006) are indicated in Table \ref{table:calibrations}, and give an idea of the uncertainties due to the SFR calibration method. 
Our sample is too incomplete to compute a best fit slope, but the method developed here could be applied to more high redshift galaxies. 

\begin{table}
\centering
\caption{Mean, median, and standard deviation of the depletion time of the clumps obtained through three different calibrations of the SFR: Kewley et al. (2004) [OII] calibration, UV+24$\mu$m calibration from Tacconi et al. (2013) and Moustakas et al. (2006) [OII] B-band calibration.}
\label{table:calibrations}
\begin{tabular}{llll}
\hline 
\hline
\multicolumn{1}{c}{SFR calibration} & \multicolumn{3}{c}{Depletion time [Gyr]}\\ \cline{2-4}
& mean & median & std. dev.\\
\hline
Kewley et al. (2004) & 1.9 & 2.0 & 1.2  \\
Tacconi et al. (2013) & 2.1 & 0.7 & 2.9\\
Moustakas et al. (2006) & 1.8& 1.6& 1.3\\
\hline
\end{tabular}
\end{table}

\section{Discussion and conclusion}

\subsection{Advantages of the method}

We have shown how the various ensembles of clumps can be separated by their kinematics in PV
diagrams, even though the angular resolution of molecular gas and SFR data was not able to separate them in integrated
intensity. As such, the KS diagram can be obtained within regions of the resolution size ($\sim$1"). 

Previous resolved KS work at high redshift was only obtained using serendipitous amplification
by gravitational lenses. 
Decarli et al. (2012) carried out one of the first spatially resolved studies in high redshift galaxies, using [NII], FIR, and CO 
observations for two gravitationally lensed $z$$\sim$3.9 galaxies. They obtain a steep relation of slope N = 1.4 $\pm 0.2$ between the dust continuum and the molecular gas surface brightness.
Strong lenses are rare, and determination of the clumps physical parameters
depend on the lensing model.
Our method is probably more appropriate for a systematic study of the star formation at high redshift, until higher resolution 
instruments resolve the clumps. 

In the absence of high resolution molecular gas data, Swinbank et al. (2012) report adaptive optics H$\alpha$ observations of eleven kpc-scale star forming regions identified in $z = 0.84-2.23$ galaxies and measure the velocity dispersion of the ionized gas $\sigma$ and the star formation surface density $\Sigma_{\mathrm{SFR}}$. 
By assuming that $\sigma$ also corresponds to the dispersion of the cold clumps and  that the clumps are 
marginally stable with a Toomre parameter $Q \simeq 1$,  they claim a correlation between the gas surface density $\Sigma_{\mathrm{gas}}$ and $\Sigma_{\mathrm{SFR}}$. 
But the method is highly indirect, relies on many assumptions, and underestimates beam-smearing effects in the determination of $\sigma$. 

\subsection{Biases and uncertainties}

The four galaxies studied here are particularly luminous and were selected from a sample (\cite{tacconi2010}) with stellar masses and star 
formation rates that are respectively higher than $3.~10^{10}~\mathrm{M}_\odot$ and $ 40~\mathrm{M}_\odot ~\mathrm{yr}^{-1}$. A high luminosity was also needed to visualize and isolate the clumps in the PV diagrams, so this method is 
intrinsically biased towards massive galaxies. 

The main uncertainties for estimating gas and SFR surface densities come from the SFR calibration and from
 the values used for $\alpha$ = $\mathrm{M}_{\mathrm{H}_2}/\mathrm{L}^\prime_{\mathrm{CO}}$, for the CO(3-2)/CO(1-0) luminosity ratio, and for the extinction 
A$_{\mathrm{H}\alpha}$.
These quantities could vary significantly from one galaxy to another and within each galaxy, thus increasing the scatter and the uncertainty of our measurements.
For example, the CO(3-2) transition is less directly related to the molecular gas mass than the CO(1-0), and variations in the CO luminosity ratios as high as of a factor $\sim$2 can be observed within a single galaxy (e.g. \cite{koda}). Significant variations in extinction values from one substructure to another  within each galaxy are also expected, as observed in local galaxies (e. g. \cite{scoville}) and high redshift galaxies. 
Genzel et al. (2013) notably uses the CO(3-2) and H$\alpha$ lines, as well as HST multiband images of a high redshift galaxy belonging to the same sample as ours to show that the methodology used to correct for extinction has a certain influence on the shape of the KS relation, especially on its slope. 
We thus expect that using a resolved extinction map instead of a single value would significantly change the depletion time we obtain. Nevertheless, identifying the ensembles of clumps in PV diagrams and the low spatial resolution of the [OII] and CO measurements prevent us from finding the corresponding structures in two-dimensional images and deriving extinction maps for our ensembles of clumps. 
As seen in section \ref{section:sfr}, the values of the SFR are highly dependent on the calibration and, since the scatter in the observed [OII]/H$\alpha$ flux ratio is always higher than 32\% (\cite{moustakas}), this remains a lower limit of the uncertainty in the SFR determined from [OII].
We  expect our final values for the mass of gas and the SFR to be determined with uncertainty factors at least as high as $2$ or $3$.

\subsection{Conclusion}

Our results, as well as most other observations (\cite{genzel}, \cite{tacconi2010}, \cite{tacconi2013}, \cite{decarli}, \cite{swinbank}), indicate that the star formation scaling law between SFR and gas surface densities is not significantly different at high redshift than in the local Universe. 
Our limited sample of $\sim$8 kpc-scale ensembles of clumps of distant galaxies is compatible with a constant depletion time of 1.9 Gyr, which is of the same order of magnitude as measurements at lower redshift. This adds to the growing evidence that the star formation processes ten billion years ago were similar to the ones that are observed in the local Universe. 
The method developed here could be applied to a more significant sample of high redshift galaxies to obtain more statistically robust results.


\begin{acknowledgements}
J. Freundlich acknowledges support by the  \'Ecole Normale sup\'erieure (ENS, Paris), and is thankful to Philippe Salom\'e for numerous technical tips and Martin Stringer for his proofreading and suggestions. The authors wish to thank the anonymous referee, whose comments have led to significant improvements in this paper.
This work is based on observations carried out with the IRAM Plateau de Bure Interferometer. IRAM is supported by INSU/CNRS (France), MPG (Germany) and IGN (Spain). 
This work also makes use of data from AEGIS, a multiwavelength sky survey conducted with the Chandra, GALEX, Hubble, Keck, CFHT, MMT, Subaru, Palomar, Spitzer, VLA, and other telescopes and supported in part by the NSF, NASA, and the STFC.
\end{acknowledgements}


\begin{thebibliography}{}
   \bibitem[Bigiel et al. 2008]{bigiel} Bigiel, F., Leroy, A., Walter, F., et al., 2008,  ApJ, 136, 2846   
   \bibitem[Bigiel et al. 2011]{bigiel2011} Bigiel, F., Leroy, A., Walter, F., et al., 2011,  ApJ, 730, L13
   \bibitem[Blanton \& Roweis 2007]{blanton} Blanton, M. R., \& Roweis, S., 2007, ApJ, 133, 734
   \bibitem[Bournaud \& Elmegreen 2009]{bournaud} Bournaud, F., \& Elmegreen, B. C., 2009, ApJ, 694, L158
   \bibitem[Coil et al. 2004]{coil} Coil, A. L., Newman, J. A., Kaiser, N., et al., 2004, ApJ, 617, 765C
   \bibitem[Cooper et al. 2011]{cooper2011} Cooper, M. C., Aird, J. A., Coil, A. L., 2011, ApJS, 193, 14C
   \bibitem[Cooper et al. 2012]{cooper} Cooper, M. C., Newman, J. A., Davis, M., et al., 2012, ASCL, 1203.003
   \bibitem[Daddi et al. 2007]{daddi} Daddi, E., Dickinson, M., Morrison, G., et al., 2007, ApJ, 670, 156
   \bibitem[Davis et al. 2003]{davis03} Davis, M., Faber, S. M., Newman, J. A., et al., 2003, Proc. SPIE, 4834, 161
   \bibitem[Davis et al. 2007]{davis07} Davis, M., Guhathakurta, P., Konidaris, N. P., et al., 2007, ApJ, 660, L1
   \bibitem[Decarli et al. 2012]{decarli} Decarli, R., Walter, F., Neri, R., et al., 2012, ApJ, 752, 2D
   \bibitem[Dekel et al. 2009]{dekel2009} Dekel, A., Birnboim, Y., Engel, G., et al., 2009, Nature, 457, 451D
   \bibitem[Dekel, Sari \& Ceverino 2009]{dekelsari} Dekel, A., Sari, R., \& Ceverino, D., 2009, ApJ, 703, 785
   \bibitem[Dickman et al. 1986]{dickman86} Dickman, R. L., Snell, R. L., \& Schoerb, F. P., 1986, ApJ, 309, 326
   \bibitem[Faber et al. 2002]{faber} Faber, S. M., Phillips, A. C., Kibrick, R. I., et al. 2002, Proc. SPIE, 4841, 1657 
   \bibitem[F\"orster Schreiber et al. 2011]{forster} F\"orster Schreiber, N. M., Shapley, A. E., Genzel, R., et al., 2011, ApJ, 739, 45F
   \bibitem[Genzel et al. 2010]{genzel} Genzel, R., Tacconi, L. J., Gracia-Carpio, J., et al., 2010, MNRAS, 407, 2091
   \bibitem[Genzel et al. 2013]{genzel2013} Genzel, R., Tacconi, L. J., Kurk, J., et al., 2013, submitted to ApJ, arXiv:1304.0668
   \bibitem[Kennicutt 1998a]{kennicutt} Kennicutt, R. C. Jr., 1998a, ApJ, 498, 541
   \bibitem[Kennicutt 1998b]{kennicutt98} Kennicutt, R. C. Jr., 1998b, ARA\&A, 36, 189
   \bibitem[Kere\v{s} et al. 2005]{keres} Kere\v{s}, D., Katz, N., Weinberg, D. H., \& Dav\'e, R., 2005, MNRAS, 363, 2K
   \bibitem[Kewley et al. 2004]{kewley} Kewley, L. J., Geller, M. J., \& Jansen, R. A., 2004, ApJ, 127, 2002
   \bibitem[Koda et al. 2013]{koda} Koda, J., Scoville, N., Hasegawa, T., et al., 2012, ApJ, 761, 41K
   \bibitem[Leroy et al. 2008]{leroy} Leroy, A. K., Walter, F., Brinks, E., et al., 2008, ApJ, 136, 2782
   \bibitem[Leroy et al. 2012]{leroy2012} Leroy, A. K., Bigiel, F., de Blok, W. J. G., et al., 2012, ApJ, 144, 3L
   \bibitem[Leroy et al. 2013]{leroy2013}   Leroy, A. K., Walter, F., Sandstrom, K., et al., 2013, arXiv:1301.2328
   \bibitem[Mannucci et al. 2010]{mannucci} Mannucci, F., Cresci, G., Maiolino, R., et al., 2010, MNRAS, 408, 2115
   \bibitem[Moustakas et al. 2006]{moustakas} Moustakas, J., Kennicutt, R. C., \& Tremonti, C., 2006, ApJ, 642, 775M
   \bibitem[Newman et al. 2012]{newman} Newman, J. A., Cooper, M. C., Davis, M., et al., 2012, submitted to ApJS, arXiv:1203.3192
   \bibitem[Noeske et al. 2007]{noeske} Noeske, K. G.,  Weiner, B. J., Faber, S., et al., 2007, ApJ, 660, L43
   \bibitem[O'Donnell 1994]{odonnell} O'Donnell, J. E., 1994, ApJ, 422, 158
   \bibitem[Renaud et al. 2012]{renaud} Renaud, F., Kralkjic, K., \& Bournaud, F., 2012, ApJ, 760, L16
   \bibitem[Saintonge et al. 2011]{saintonge2011} Saintonge, A., Kauffmann, G., Wang, J., et al., 2011, MNRAS, 415, 61S
   \bibitem[Schlegel, Finkbeiner \& Davis 1998]{schlegel} Schlegel, D., Finkbeiner, D., \& Davis, M., 1998, ApJ, 121, 161
   \bibitem[Scoville et al. 2001]{scoville} Scoville, N. Z., Polletta, M., Ewald, S., et al., 2001, ApJ, 122, 3017
   \bibitem[Solomon et al. 1997]{solomon97} Solomon, P. M., Downes, D., Radford, S. J. E., et al., 1997, ApJ, 478, 144
   \bibitem[Solomon et al. 1987]{solomon87} Solomon, P. M., Rivolo, A. R., Barrett, J., et al., 1987, ApJ, 319, 730
   \bibitem[Swinbank et al. 2012]{swinbank} Swinbank, A. M., Smail, I., Sobral, D., et al., 2012, ApJ, 760, 130S
   \bibitem[Tacconi et al. 2010]{tacconi2010} Tacconi, L. J., Genzel, R., Neri, R., et al., 2010, Nature, 463, 781T
   \bibitem[Tacconi et al. 2013]{tacconi2013} Tacconi, L. J., Neri, R., Genzel, R., et al., 2013, submitted to ApJ, arXiv:1211.5743
   \bibitem[Van de Voort et al. 2011]{vandevoort} Van de Voort, F., Schaye, J., Booth, C. M., et al., 2011, MNRAS, 414, 2458
   \bibitem[Wyder et al. 2009]{wyder} Wyder, T. K., Martin, D. C., Barlow, T. A., et al., 2009, ApJ, 696, 1834W
   \bibitem[Wuyts et al. 2011a]{wuyts} Wuyts, S., F\"orster Schreiber, N., Lutz, D., et al., 2011, ApJ, 738, 106
   \bibitem[Wuyts et al. 2011b]{wuyts2011b} Wuyts, S., F\"orster Schreiber, N. M., van der Wel, A., et al., 2011, ApJ, 742, 96W
\end{thebibliography}
\end{document}